# Technical assessment of a novel vertical CT system for upright radiotherapy simulation and treatment planning

Jordan M. Slagowski,[1,2] Yuhao Yan,[1,2] Jessica R. Miller,[1,2] John W. Hayes,[1,2] Carson A. Hoffman,[3] Minglei Kang,[1] Carri K. Glide-Hurst[1,2]

1. Department of Human Oncology, University of Wisconsin – Madison, Wisconsin, USA
2. Department of Medical Physics, University of Wisconsin – Madison, Wisconsin, USA
3. Leo Cancer Care, Inc., Middleton, Wisconsin, USA

Corresponding authors:
Jordan M. Slagowski, PhD
Email: slagowski@wisc.edu

Carri K. Glide-Hurst, PhD
Email: glidehurst@humonc.wisc.edu

**ABSTRACT**

**Background**: Upright patient positioning for radiotherapy may offer anatomical and dosimetric advantages, improved patient comfort, and more cost-effective proton therapy by replacing large rotating gantries with a fixed beamline and rotating patient positioner. However, clinical adoption has been limited by the lack of available vertical CT systems needed to support treatment planning and image guidance.

**Purpose**: To present the first technical characterization of image quality, imaging dose, and dose calculation accuracy for a novel upright CT imaging system designed for radiotherapy applications.

**Methods**: An upright CT scanner equipped with a six-degree-of-freedom patient positioning system (Marie from Leo Cancer Care) was evaluated for radiotherapy simulation and treatment planning. Imaging dose ($CTDI_{vol}$) was measured in 16- and 32 cm diameter CTDI phantoms at 120 kVp and 200 mAs. Image quality was evaluated using an ACR-464 phantom against diagnostic CT accreditation standards. CT number accuracy was assessed by measuring the mean CT numbers within five inserts of known material. Uniformity was assessed as the difference in mean image values at the center and periphery of 20-cm and 48-cm diameter phantoms. High-contrast spatial resolution was assessed in terms of visible line pairs and a modulation transfer function (MTF). Low-contrast performance was quantified in terms of the contrast-to-noise ratio (CNR). Spatial integrity was evaluated by measuring the distance between two high-contrast fiducials separated by 100 mm. Hounsfield unit look-up tables for mass density and stopping-power-ratio were generated using an advanced electron density CT phantom. Upright CT number linearity was assessed by computing the coefficient of determination ($R^2$) versus a recumbent CT scanner. Proton and photon treatment plans were optimized on upright CT scans of an anthropomorphic thorax phantom in heterogeneous (lung, spine) and homogeneous (liver) tissue regions. Dose was forward computed on a registered CT scan acquired on a conventional recumbent CT system. Dosimetric agreement between upright and recumbent CT-based plans was evaluated using 3D global gamma analysis.

**Results**: CT imaging dose ($CTDI_{vol}$) was 23.5 mGy for the 16 cm head phantom and 10.1 mGy for the 32 cm body phantom. Mean CT numbers (HU) were within the expected range for water (1.7) and acrylic (120.8). CT numbers were slightly (5-27 HU) out-of-range for air (-950.4), polyethylene (-78.8), and bone (823.0). Image uniformity was 20.2 HU and 35.0 HU for the 20 cm and 48 cm diameter phantoms, respectively. Eight high-contrast line pairs were visualized. The MTF equaled 4.4 $cm^{-1}$ at 50% and 7.1 $cm^{-1}$ at 10%. The median CNR was 0.93, slightly below the ≥1.0 tolerance. Spatial integrity was ≤0.36 mm. Upright CT numbers trended linearly versus the reference CT scanner with an $R^2$ value of 0.9997. Gamma pass rates were ≥99.8% for photon and ≥90.6% for proton plans with 1%/1mm criteria, and ≥98.0% for all plans with 3%/2mm criteria. Dose differences were greatest for proton plans in the low-density lung region.

**Conclusion**: Upright CT image quality and dose calculation accuracy were systematically evaluated and found to be acceptable for photon and proton radiotherapy simulation and treatment planning.

## 1. INTRODUCTION

For more than half a century, nearly all radiation therapy (RT) has been delivered with patients in a recumbent position using a rotating gantry to deliver treatment beams from multiple angles.[1] The recumbent treatment position partially arose out of necessity to accommodate the horizontal geometry of CT scanners essential for radiation treatment planning. However, emerging evidence suggests that upright patient positioning (e.g. seated, perched, or standing) could provide anatomical and economic advantages while also improving patient comfort.[1,2] By replacing large (80-120 ton) rotating gantries, prone to mechanical failures, with a fixed proton beamline and an upright rotating patient positioner, treatment facilities can significantly decrease initial construction, radiation shielding, and ongoing maintenance costs, to make dosimetrically superior charged particle therapies more accessible and economically viable.[3,4]

Upright patient positioning may alleviate excessive saliva accumulation, swallowing difficulties, and dyspnea in head and neck cancer patients.[5] Thoracic cancer patients may benefit from increased lung volumes and reduced tumor motion relative to supine positioning to mitigate toxicities such as pneumonitis.[6–9] Additionally, observed shifts in organ positions, including an increased distance between the prostate apex and penile bulb,[10] a more inferior kidney location,[11] and variations in the spleen, pancreas, and bowel between upright and supine orientations, may offer dosimetric advantages. Upright positioning also offers a valuable alternative for patients who cannot lie flat due to breathing difficulties, morbid obesity, or severe pain.[1,5,12]

To reap the potential benefits of upright RT, isocentric upright patient positioning devices have been under development for decades. Motivated to minimize mediastinal tumor spread to reduce lung dose, Miller et al. introduced the first isocentric chair compatible with existing radiation treatment units.[13] Investigators at the University of Texas MD Anderson Cancer Center developed an upright positioner made of acrylic and wood that attaches to the treatment couch of a linear accelerator.[5] Since then, other innovative designs have emerged including validated hexapod platforms,[14,15] a portable rotating platform compatible with existing proton therapy machines,[16] and an upright patient positioning system mounted to a six-degree-of-freedom robotic arm.[17] Two recent review papers comprehensively describe the evolution and design of existing patient positioners.[18,19]

Despite the successful development of upright patient positioners, the anticipated benefits of upright RT have yet to be fully realized, largely due to the lack of accessible vertical CT imaging systems necessary for anatomical assessment, treatment planning, and patient alignment. Recently, an upright helical CT system was installed for diagnostic radiology applications at Keio University in collaboration with Canon Medical Systems (Otawara, Japan).[20] The 320 row-detector system has provided critical data distinguishing anatomical and physiological differences between upright and supine positioning.[8,21–23] Two vertical CT systems integrated with upright patient positioners for radiotherapy applications are now commercially available. P-Cure Ltd. (Shilat, Israel) developed the Patient Robotic Positioning and Imaging System (P-ARTIS) which combines a chair mounted to a six-degree-of-freedom robotic arm with a vertical 4DCT imaging system.[17] The CT system uses a Phillips Brilliance Big Bore scanner with modified motion control.[17] As a result, standard clinical features including respiratory motion management are retained. The P-Cure system is currently in clinical use at Northwestern Medicine Proton Center (Chicago, Il, USA) and Hadassah Medical Center (Hadassah, Israel).[9,17,24] Leo Cancer Care, Inc. (Middleton, WI, USA) has developed Marie, designed specifically for upright radiation therapy, which combines a patient positioner fully integrated with a vertical fan beam CT scanner.[25] To the best of our knowledge, system performance has not yet been characterized and reported for the Leo Cancer Care CT system.

The purpose of this study was to perform a technical characterization of a novel vertical CT scanner, the Marie System, for applications in photon and proton radiotherapy. An overview of the upright CT system, image acquisition protocols, and measurements of imaging dose are provided. Image quality was assessed versus American College of Radiology (ACR) accreditation standards in terms of low- and high-contrast resolution, uniformity, spatial integrity, and CT number accuracy. CT-number to mass-density and stopping-power-ratio (SPR) calibrations were performed.  Photon and proton dosimetric calculations were computed on upright CT images and comparisons were made to standard of care conventional (e.g. recumbent) CT data as a first step toward clinical implementation.

## 2. METHODS

### 2.1 Upright CT system overview

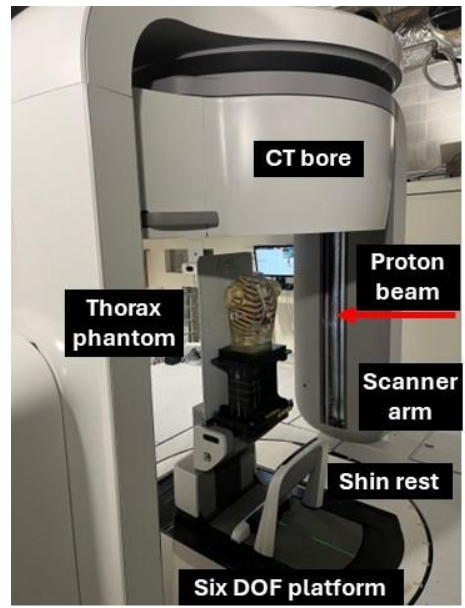

**Figure 1:** Upright CT system for radiation therapy simulation and treatment planning.

The Marie system (Leo Cancer Care, Middleton, WI, USA), shown in Figure 1, integrates a helical CT scanner with an upright patient positioner and was developed for gantry-less, image-guided radiation therapy and simulation. The CT bore is mounted to two scanner arms containing counterweights that enable upright patient-imaging over a 130 cm vertical range with tilt angles up to +/-15°.[25] The large CT bore and field-of-view (FOV) allow the system to image patients with large body habitus and immobilization devices without truncation as required for treatment planning. This study evaluated the serial number 6 system installed at the University of Wisconsin – Madison in 2024. System parameters are summarized in Table 1.

The CT system contains a rotating anode x-ray tube from Varex Imaging (Salt Lake City, UT, USA) and bowtie filter. The 32-slice detector uses X-Tile elements from Detection Technology (Billerica, MA, USA) with ultrafast ceramic gadolinium oxysulfide scintillators coupled with back-side illuminated photodiodes. [25]
Image reconstruction is performed by filtered backprojection. Available filters include the Ram-Lak to preserve high-frequency detail and Shepp-Logan that applies a low-pass filter to reduce noise.

**Table 1:** Upright CT system x-ray tube, detector, and reconstruction parameters.

| CT System Parameters | |
| --- | --- |
| CT bore diameter | 85 cm |
| Vertical travel range | 130 cm |
| X-ray tube kVp | 120 |
| mA | 10-250 |
| Rotation time (360°) | 1.0 s |
| Scintillator type | Gadolinium Oxysulfide (GOS) |
| Beam width at isocenter | 19.3 mm |
| Reconstruction FOV | 16-62.3 cm |
| Reconstruction algorithm | Filtered Backprojection |
| Reconstruction filters | Ram-Lak, Shepp-Logan |
| Stored bit depth | 16-bit |

## 2.2 CT imaging protocols

All CT imaging was performed at 120 kVp in this study with acquisition parameters summarized in Table 2. The *Thorax Vendor Default* CT imaging protocol has a 57.0 cm FOV and 2.0 mm slice thickness. An additional protocol with a 50.0 cm FOV (*Thorax UW)* was created and used for all treatment planning studies in this work. A custom high-resolution protocol (*ACR QA*) with a 21.0 cm FOV and 1.0 mm slice thickness was created for CT image quality evaluation. Automatic exposure control (AEC) was set off for all protocols.

**Table 2:** Upright CT imaging protocols.

| Protocol | Energy [kV] | Exposure [mA][a] | Rotation Time [s] | Pitch | Filter | FOV Diameter [cm] | Slice Thickness [mm] |
|---|---|---|---|---|---|---|---|
| Thorax Vendor Default | 120 | 250 | 1.0 | 1.0 | Ram-Lak | 57.0 | 2.0 |
| Thorax UW | 120 | 250 | 1.0 | 1.0 | Ram-Lak | 50.0 | 2.0 |
| Head Hi-Res Adult | 120 | 250 | 1.0 | 1.0 | Ram-Lak | 30.0 | 1.0 |
| ACR QA | 120 | 250 | 1.0 | 1.0 | Ram-Lak | 21.0 | 1.0 |

a) AEC turned off

## 2.3 CT imaging dose

Upright CT imaging dose was measured in terms of $CTDI_{vol}$ using a 100-mm-long pencil ionization chamber (Accu-Dose+, Radcal, Monrovia, CA, USA) with an accredited calibration for a 16 cm diameter CTDI head phantom and a 32 cm diameter body phantom. CT imaging was performed at 120 kVp, 200 mA, 32 x 0.605 mm collimation width, and a 1.0 s rotation time in an engineering enabled axial scan mode with AEC turned off. For all measurements, the CT ring was placed at the top of the scanner arms in the park position. For each phantom, a total of three measurements were performed and averaged for the center hole ($D_c$) and each of the four peripheral ($D_p$) holes. The weighted CTDI ($CTDI_w$) was computed according to Equation 1.

$$CTDI_w = \frac{1}{3} \cdot D_c + \frac{2}{3} \cdot D_p \quad (1)$$

Weighted CTDI was converted to $CTDI_{vol}$ according to Equation 2 with a pitch value of 1.0 for all imaging protocols.

$$CTDI_{vol} = \frac{1}{pitch} \cdot CTDI_w \quad (2)$$

To assess suitability for human subject imaging, $CTDI_{vol}$ values for the *Head Hi-Res Adult* and *Thorax Vendor Default* protocols were compared to dose values from several published recumbent CT simulation protocols.[26,27]

## 2.4 CT image quality evaluation

Upright CT image quality was assessed following ACR guidelines for diagnostic CT quality control and scanner accreditation.[28] ACR guidelines were chosen due to their well established procedures and defined tolerances. The 20 cm diameter by 16 cm tall ACR Model 464 CT accreditation phantom (Sun Nuclear, Melbourne, FL, USA) was scanned using the *ACR QA* protocol summarized in Table 2. The phantom was scanned on seven separate instances over a 54-day period. Image quality was assessed in terms of low-contrast performance, CT number accuracy, image uniformity, and high-contrast spatial resolution, following ACR specified procedures.[28] Image quality tolerances are specified in Table 4 and Table 5. Results are reported as the median, minimum, and maximum measured values for each image quality metric.

*2.4.1 Low-contrast performance*
Low-contrast performance was measured in terms of the contrast-to-noise ratio (CNR). The mean image value ($\mu_{target}$) was computed within a 100 mm$^2$ circular region-of-interest (ROI) placed within the 25 mm diameter target of Module 2 of the phantom. The mean ($\mu_{background}$) and standard deviation ($\sigma_{background}$) of image values in an adjacent background ROI were measured. CNR was computed as: $CNR = (\mu_{target} - \mu_{background})/\sigma_{background}$.

*2.4.2 CT number accuracy*
CT number accuracy was evaluated for five material inserts including air, polyethylene, water, acrylic, and Teflon (bone equivalent). Mean CT numbers were computed within 20.0 mm diameter circular ROIs (80% of the insert size) as delineated in Figure 2C.

*2.4.3 Image uniformity*
Image uniformity was determined by computing the mean image values in five ROIs placed at the center and periphery of module 3 of the ACR CT phantom as shown in Figure 2B. Uniformity was quantified as the difference between the center ROI and the mean of the peripheral ROIs.[28] In addition to measurements within the 20 cm diameter ACR-464 phantom, a 48 cm diameter Helios IQ Cal 48 Poly phantom (Model 2144721-2, GE Healthcare) was also imaged to evaluate uniformity over a larger FOV.

*2.4.4 High-contrast spatial resolution*
To quantitively assess high contrast spatial resolution, the radial modulation transfer function (MTF) was derived from the uniformity module of the ACR-464 phantom.[29] The MTF was computed as the fast Fourier transform of a line spread function obtained from differentiating the edge spread function measured from oversampling the air and phantom interface. Spatial frequencies (cm$^{-1}$) corresponding to MTF values of 0.5 and 0.1 were reported to characterize the maximum detectable spatial frequency. Qualitatively, the maximum number of visible line pairs per cm (lp/cm) was recorded by a qualified medical physicist after optimizing the window and level of the phantom image for line pair visualization.

*2.4.5 Spatial integrity*
Spatial integrity was assessed by measuring the distance between the centers-of-mass of two high contrast fiducial markers located in the uniformity module of the ACR-464 phantom.[30,31] The measured distance was compared versus the nominal physical distance of 100 mm.

**2.5 CT number to density and stopping-power-ratio**
CT number to mass density and SPR calibrations were performed to enable photon and proton dose calculations, respectively, within the RayStation 2024A SP3 (RaySearch Laboratories AB, Stockholm, Sweden) treatment planning system (TPS). A 40 cm by 30 cm Gammex Advanced Electron Density Phantom (Gammex, Middleton, WI, USA) was scanned using the upright Marie CT scanner with the *Thorax UW* protocol (Table 2) and a conventional SOMATOM Definition CT (Siemens Healthineers, Erlangen, Germany) scanner at matched tube voltage and mAs. The phantom was scanned for five different configurations of water and bone inserts following consensus guidelines designed to reduce inter-center variation in SPR determination.[32] Mean CT numbers were extracted from cylindrical ROIs within the inserted rods using an in-house version of Pylinac modified to average CT values over 10 image slices.[33] A Hounsfield unit look-up table (HLUT) to SPR was then generated using the Bethe equation with ICRU-49 mean excitation energies and software provided by the consensus guidelines (https://github.com/CTinRT/HLUT-guide).[32] Upright CT numbers, along with their corresponding mass densities and SPR values, were input into the RayStation TPS and compared with those from the recumbent CT simulator. CT number linearity was assessed by computing the coefficient of determination ($R^2$) from a linear regression of upright versus recumbent CT numbers (HU) at matched insert densities.

**2.6 Photon and proton dose calculations**

The suitability of upright CT for photon and proton treatment planning was evaluated using an anthropomorphic adult male thorax phantom (RSD-111T, Radiology Support Devices, USA). The rigid phantom maintains consistent internal anatomy between upright and recumbent orientations and contains realistic internal features including lung and bone heterogeneities. The phantom was scanned on the recumbent SOMATOM Definition Edge scanner and the upright Marie CT scanner as shown in Figure 1 after aligning the phantom at the CT bore isocenter using lasers. The *Thorax UW* protocol (Table 2) was used for upright scanning. For the recumbent scan, an institutional chest CT protocol was used with acquisition parameters including the reconstruction FOV (50.0 cm), voxel dimensions (0.98 mm) and slice thickness (2.0 mm) adjusted to match the upright CT protocol.

Planning target volumes (PTV) mimicking a lung tumor (73.4 $cm^3$), a spinal bone metastasis (23.2 $cm^3$), and a liver tumor (17.5 $cm^3$), were contoured on the upright CT image volume. These disease sites were selected to mimic radiation treatments in heterogenous (lung, bone) and homogenous (liver) anatomical regions. Organs-at-risk including the lungs, heart, esophagus, and spinal cord were also delineated. The recumbent CT dataset was rigidly registered to the upright CT scan focusing on the vertebral bones and a copy of the contours was transferred from the upright CT to the recumbent CT images.

Radiation treatment plans were optimized for each PTV on the upright CT images using the RayStation TPS. Prescription doses and planning objectives for targets and organs-at-risk are summarized in Table 3.[34–37] For photon planning, a VMAT technique was used for lung and spine treatments, while a 3D conformal method was used for the liver treatment, using a commissioned Varian TrueBeam beam model. Photon dose calculations were performed using the collapsed cone v5.9 algorithm. All proton plans used Hitachi PROBEAT pencil beam scanning. Robust optimization was performed with +/-3.5% range uncertainty and +/-5 mm setup uncertainties. Proton dose calculations were performed using Monte Carlo v5.6 with 0.1% statistical uncertainty following standard clinical protocols. After optimizing each plan on the upright CT, the 3D dose distribution was re-computed on the rigidly registered recumbent CT using the appropriate scanner specific HLUT. All other parameters were held fixed to quantify the dosimetric agreement between upright and recumbent CT images. A 2.0 mm isotropic dose grid was used for all dose calculations.

**Table 3:** Treatment planning goals.

| Target or Organ-at-Risk | Planning Goal |
| --- | --- |
| Spine PTV | D95% > 27 Gy (3 fx) |
| Lung | V11.6 Gy < 1500 $cm^3$ |
| Lung | V12.4 Gy < 1000 $cm^3$ |
| Spinal Cord | D0.03 $cm^3$ < 18 Gy |
| Lung PTV | D95% > 74 Gy (37 fx) |
| Esophagus | D0.03 $cm^3$ < 80 Gy |
| Spinal Cord | D0.03 $cm^3$ < 45 Gy |
| Heart | Mean < 26 Gy |
| Lung | Mean < 20 Gy |
| Lung | V20 Gy < 35% |
| Lung | V10 Gy < 45% |
| Lung | V5 Gy < 65% |
| Liver PTV | D95% > 50 Gy (5 fx) |
| Spinal Cord | D0.03 $cm^3$ < 22 Gy |

Dosimetric agreement between recumbent and upright CT imaging was quantified in terms of voxel-by-voxel percent differences and 3D gamma analysis.[38] The open source PyMedPhys software package was used to perform gamma analysis.[39] To highlight differences, gamma pass rates were reported with stringent 1 mm distance-to-agreement and 1% dose difference criteria (1%/1mm) with global normalization, 2%/2mm, and 3%/2mm following TG-218 recommended criteria.[40] A 10% dose threshold was used for all scenarios.

## 3. RESULTS
### 3.1 CT imaging dose
The measured $CTDI_{vol}$ was 23.5 mGy for the 16 cm head phantom and 10.1 mGy for the 32 cm body phantom. For the *Head Hi-Res Adult* protocol (Table 2), $CTDI_{vol}$ equaled 29.4 mGy. $CTDI_{vol}$ was 12.6 mGy for the *Thorax Vendor Default* protocol.

### 3.2 CT image quality evaluation
Upright CT images of the ACR-464 phantom are presented in Figure 2 for each module with window and level settings appropriate to the specific image quality task. Objects of interest including the 25-mm diameter low contrast rod (Figure 2A), CT number inserts (Figure 2C) and high-contrast line-pairs (Figure 2D) are identifiable. Quantitative image quality metrics are summarized in Table 4 and Table 5.

*3.2.1 Low-contrast performance*

Low-contrast performance in terms of CNR is summarized in Table 4. CNR ranged from 0.81 to 1.10 with a median value of 0.93, slightly below the ≥1.0 tolerance. Qualitatively, the 25 mm rod was visible (Figure 2A). None of the 6 mm rods could be visualized regardless of the window and level setting for low-contrast detectability.

*3.2.2 CT number accuracy*

The mean and standard deviation of the measured CT numbers for each of the ACR 464 material inserts are summarized in Table 5. The ACR specified tolerance range is provided. Mean CT numbers were within the expected range for water and acrylic. CT numbers were slightly (5-27 HU) out-of-range for air, polyethylene, and bone.

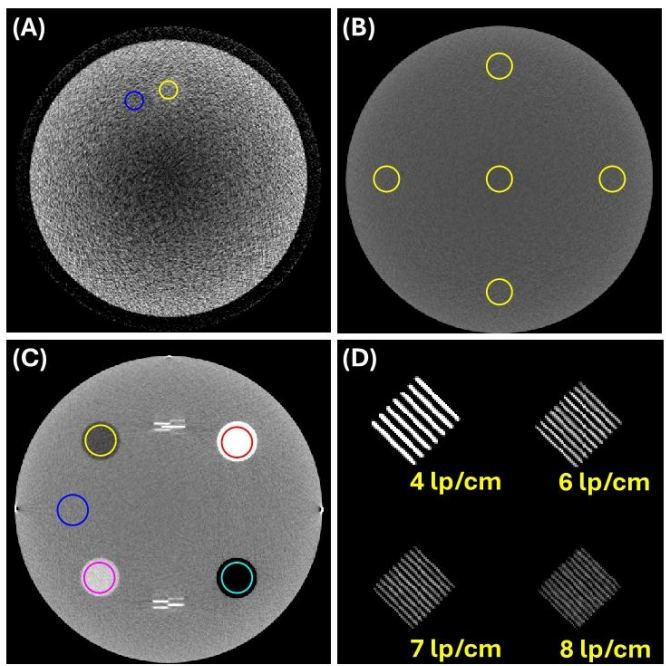

**Figure 2**: ACR head phantom images acquired on the upright CT scanner are shown for the (A) low contrast resolution module [50 HU, 150 HU], (B) uniformity module [-200 HU, 400 HU], (C) HU accuracy module [-200 HU, 200 HU] and (D) zoomed regions-of-interest of the high contrast spatial resolution module [900 HU, 1300 HU].

**Table 4:** Upright CT image quality metrics.

| Image Quality Metric | Median [Min, Max] | ACR Tolerance |
|---|---|---|
| Uniformity (HU) | 20.2 [18.9, 20.3] | ≤7.0 |
| Low contrast resolution (CNR) | 0.93 [0.81, 1.10] | ≥1.0 |
| Spatial Integrity (mm) | 0.23 [0.15, 0.36] | ≤1.0 mm[a] |
| Spatial resolution (lp/cm, 50% MTF) | 4.4 [3.8, 4.7] | N/A |
| Spatial resolution (lp/cm, 10% MTF) | 7.1 [6.6, 7.2] | N/A |
| Spatial resolution (lp/cm, qualitative) | 8.0 [8.0, 8.0] | ≥6 |

a) Tolerance per AAPM TG-66.[31]

**Table 5:** Upright CT number accuracy.

| Material | Median [Min, Max] CT Number (HU) | Standard Deviation (HU) | ACR Specified Range (HU) |
|---|---|---|---|
| Air | -950.4 [-952.6, -949.5] | 12.1 [11.4, 13.3] | [-1005, -970] |
| Polyethylene | -78.8 [-83.3, -75.9] | 11.8 [10.6, 12.7] | [-107, -84] |
| Water | 1.7 [-0.2, 5.4] | 12.0 [11.1, 12.3] | [-7, 7] |
| Acrylic | 120.8 [119.6, 124.0] | 12.0 [10.6, 12.7] | [110, 135] |
| Bone | 823.0 [819.5, 830.2] | 12.9 [11.4, 14.0] | [850, 970] |

*3.2.3 Image uniformity*

Uniformity assessed using the ACR-464 phantom was 20.2 HU (Table 4, median) which exceeded the ACR specified tolerance of ≤7.0 HU. No ring artifacts were observed. Uniformity assessed in the 48 cm diameter Helios IQ Cal 48 Poly phantom (Figure 3) equaled 35.0 HU.

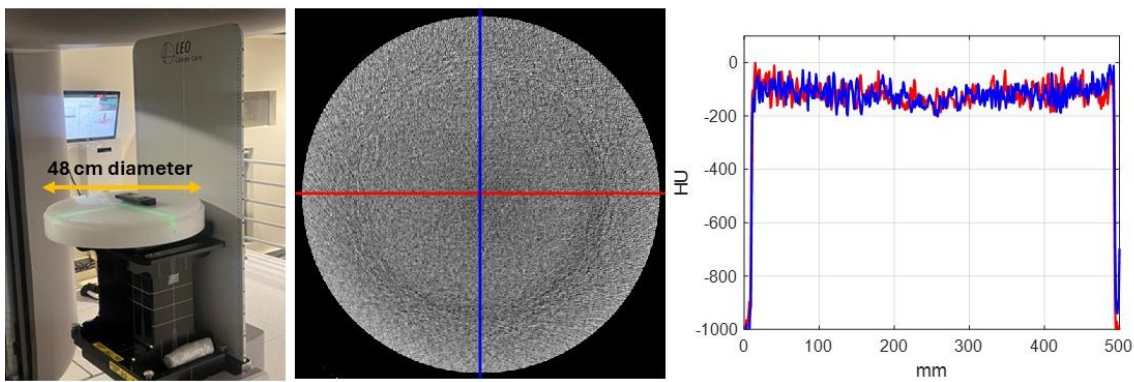

**Figure 3:** An upright CT image [-300 HU, 100 HU] of a 48 cm diameter Helios IQ Cal phantom is shown with horizontal and vertical line profiles.

### 3.2.4 High-contrast spatial resolution

Figure 2D presents magnified images of a subset of the high-contrast line-pairs within the phantom. The maximum number of line-pairs visualized was 8 line-pairs per cm (lp/cm) which satisfied the ACR tolerance of ≥6 lp/cm. Figure 4 shows the MTF. The MTF was reduced to 4.4 cm$^{-1}$ at 0.5 and 7.1 cm$^{-1}$ at 0.1 (Table 4, median values). The number of line pairs qualitatively observed in Figure 2D were consistent with the MTF measurements.

### 3.2.5 Spatial integrity

The median geometric error equaled 0.23 mm (Table 4). All measured errors were less than or equal to 0.36 mm thereby satisfying the AAPM Task Group 66 tolerance of ≤1 mm.[31]

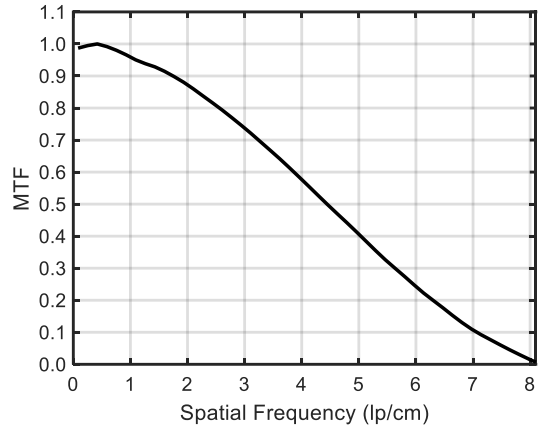

**Figure 4:** Radial modulation transfer function for the ACR QA CT imaging protocol.

### 3.3 CT number to density and stopping-power-ratio

An example upright CT image of the Advanced Electron Density Phantom is presented in Figure 5A. CT number linearity assessed versus the reference CT simulator is presented in Figure 5B demonstrating excellent agreement with the reference CT scanner ($R^2$ = 0.9997). Qualitatively, the upright CT number to mass density (Figure 5C) and SPR (Figure 5D) calibrations show close agreement with the recumbent CT simulator; the mean absolute HU difference was 28 HU, and the largest discrepancy was 52 HU which occurred in the bone region (1.93 g/cm$^3$).

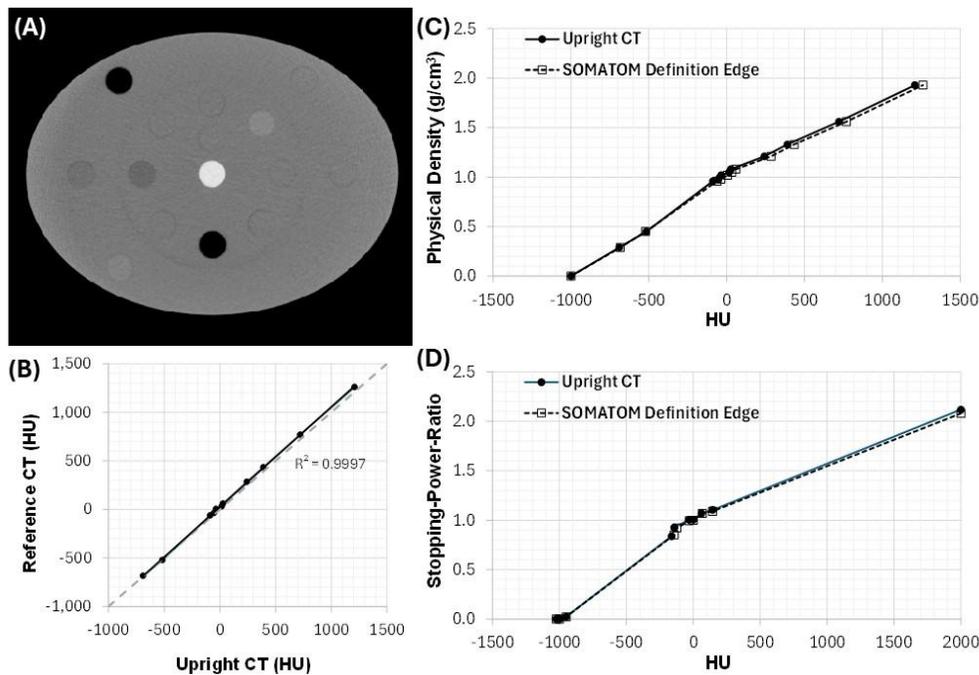

**Figure 5:** (A) An upright CT image of the CT number to electron density phantom is shown [WL = 0 HU, WW = 1000 HU]. (B) Upright CT number linearity is evaluated versus a reference CT scanner ($R^2$=0.9997). CT number versus mass density (C) and SPR (D) are shown for the upright CT scanner versus a recumbent CT system.

### 3.4 Photon and proton dose calculations

To demonstrate image quality under clinically relevant display settings, Figure 6 presents the optimized liver, lung, and spine proton dose distributions on upright and recumbent CT images in coronal and sagittal views of the thorax phantom. Axial dose distributions are shown for protons in Figure 7 and photons in Figure 8. Gamma analysis and percentage dose difference maps accompany each planning scenario to highlight regions of greatest dosimetric disagreement between upright and recumbent CT based dose distributions.

Gamma analysis pass rates are summarized in Table 6. Using a 1%/1mm criteria, gamma pass rates were ≥99.8% for all photon treatment plans, demonstrating excellent agreement between dose calculations on upright and recumbent CT datasets. For proton plans, gamma pass rates were ≥97.0% for the spine and liver plans but were slightly lower at 90.6% for the lung plan. Examination of the percent difference and gamma analysis maps (Figure 7) reveals the region with the lowest pass rate corresponds to the low-density lung region adjacent to the high dose, high gradient PTV region. Increasing the gamma criteria to 2%/2mm improved the pass rate to 97.7%. All photon and proton plans satisfied the 95% pass rate tolerance recommended by AAPM TG-218 (3%/2mm criteria).[40] Dosimetric consistency between upright and recumbent CT volumes is further supported by Figure 9 which presents dose volume histograms for a composite (i.e. summed) dose distribution of the spine, liver, and lung plans. The dose volume histograms show close agreement with target coverage differences of less than 0.9%. All organ-at-risk planning objectives listed in Table 3 were met, with maximum differences under 2.9%, except for the mean proton heart dose, which differed by -23.6%. This large percentage difference is attributed to the low absolute doses associated with proton planning which were 0.68 Gy and 0.89 Gy for the upright and recumbent CT-based plans, respectively.

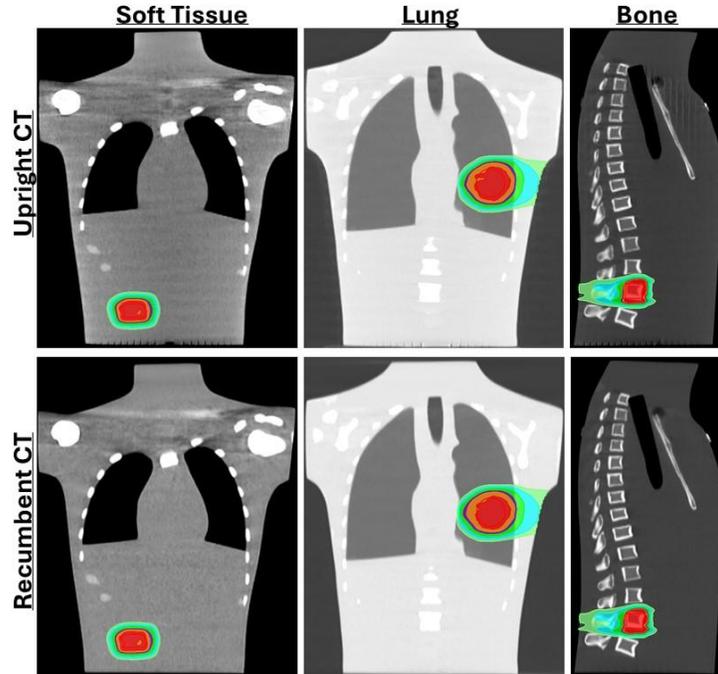

**Figure 6:** Upright and recumbent CT images of an anthropomorphic thorax phantom with proton doses are presented for soft tissue [40, 350] HU, lung [-600, 1600] HU, and bone [450, 1600] HU windows and levels.

**Table 6:** Global gamma analysis pass rates are presented for photon and proton dose calculations performed on upright versus recumbent CT acquired images.

|  | Photon | | | Proton | | |
| --- | --- | --- | --- | --- | --- | --- |
|  | 1%/1mm | 2%/2mm | 3%/2mm | 1%/1mm | 2%/2mm | 3%/2mm |
| Liver | 100.0% | 100.0% | 100.0% | 99.9% | 100.0% | 100.0% |
| Spine | 99.8% | 100.0% | 100.0% | 97.0% | 99.5% | 99.7% |
| Lung | 100.0% | 100.0% | 100.0% | 90.6% | 97.7% | 98.0% |

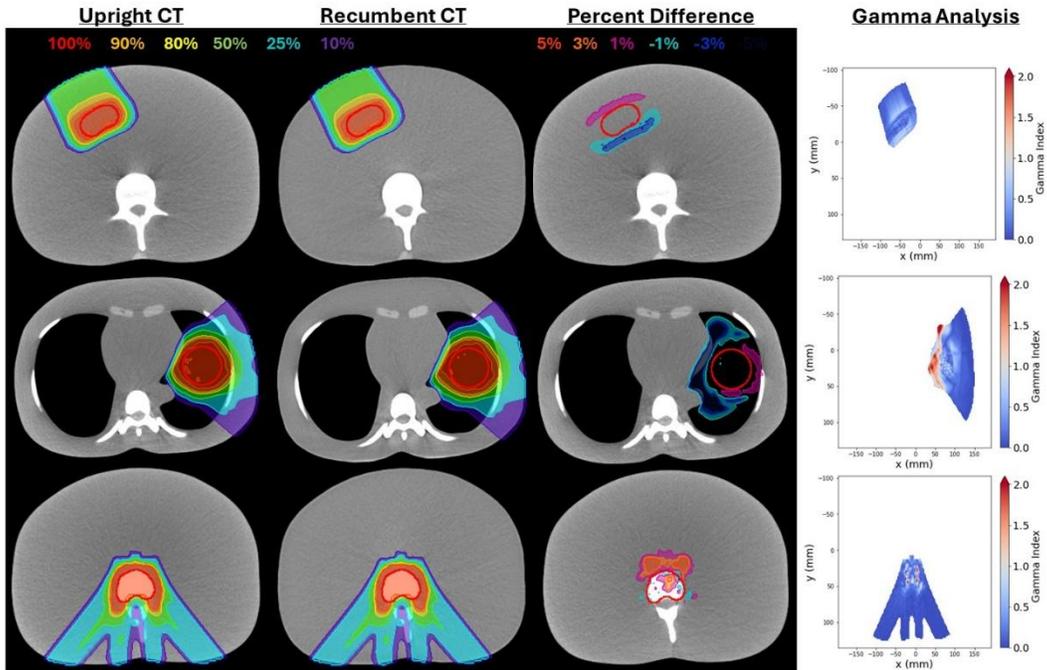

**Figure 8:** Proton liver (top row), lung (middle row), and spine (bottom row) plans optimized and calculated on upright CT images are presented (first column) versus dose re-computed on registered recumbent CT images (second column). Percent dose difference (third column) and gamma analysis maps (fourth column) highlight local differences.

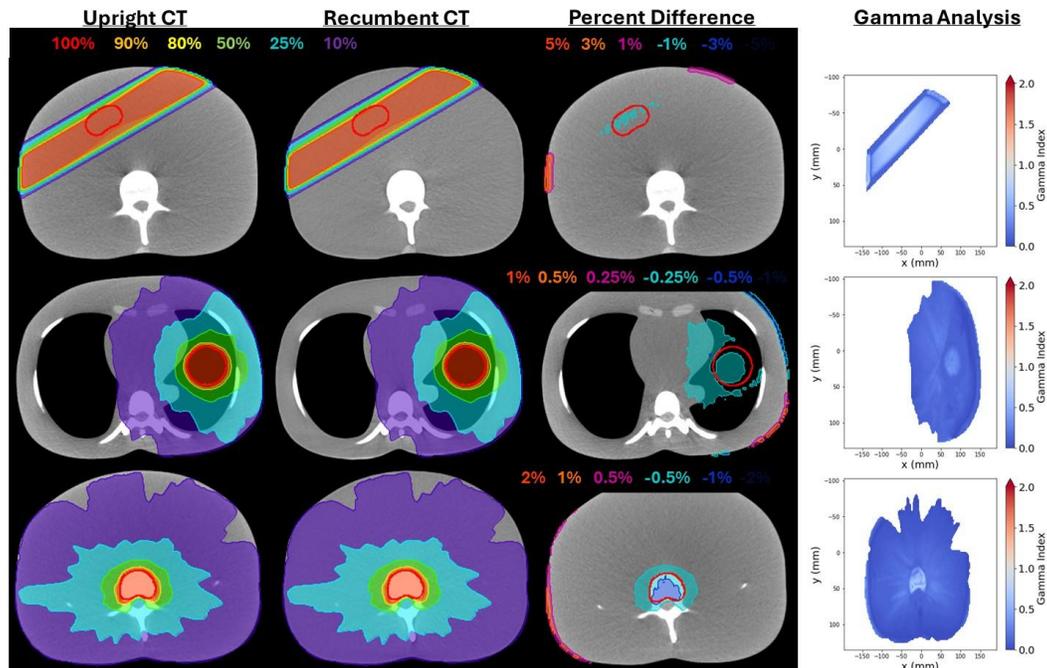

**Figure 7:** Photon liver (top row), lung (middle row), and spine (bottom row) plans optimized and calculated on upright CT images are presented (first column) versus dose re-computed on registered recumbent CT images (second column). Percent dose difference (third column) and gamma analysis maps (fourth column) highlight local differences.

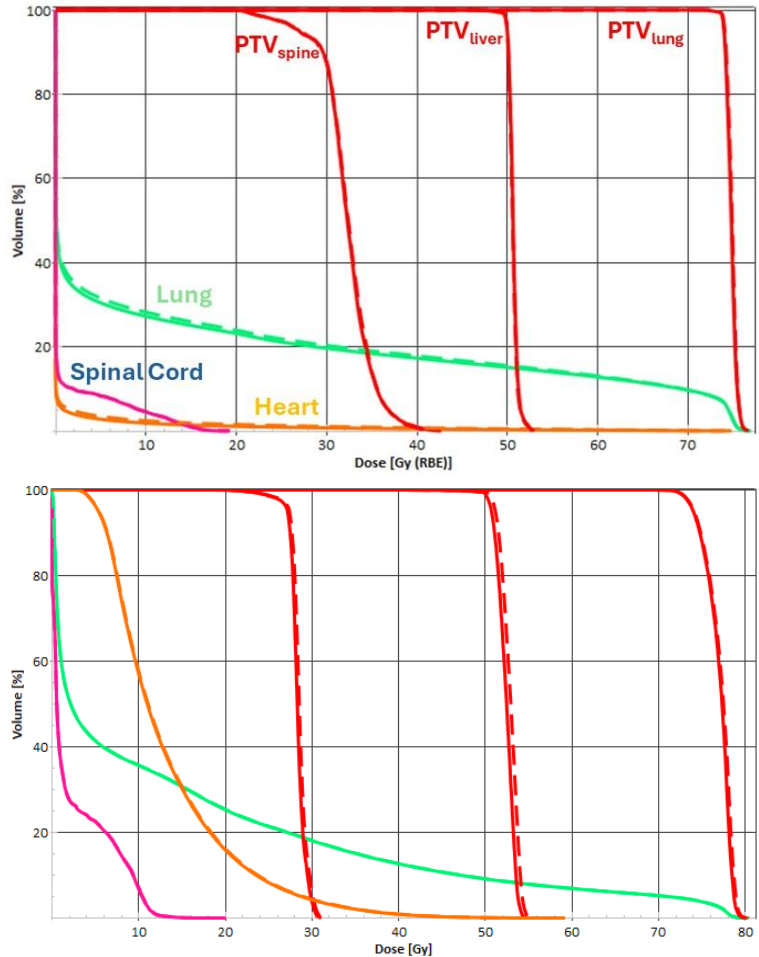

**Figure 9:** Dose volume histograms for summed spine, liver, and lung plans are shown for protons (top) and photons (bottom). Solid lines represent dose computed on upright CT images; dashed lines represent dose on recumbent CT images.

## 4. DISCUSSION

This work presents the first CT images acquired on a novel upright CT simulator and demonstrates the first photon and proton dose calculation results on these datasets. Our findings show that the system achieves clinically acceptable image quality and dose calculation accuracy supporting its integration into radiotherapy simulation and treatment planning workflows. By coupling an online helical CT scanner with a six degree-of-freedom upright patient positioner, the Marie system eliminates the need for large, rotating gantries, which may reduce treatment room size and improve accessibility to proton therapy.

Measured $CTDI_{vol}$ results were comparable to those reported in the literature for recumbent CT simulators. For example, the *Head Hi-Res Adult* $CTDI_{vol}$ was 29.4 mGy which is within the 50[th]-75[th] percentile range (20-33 mGy) of head and neck CT planning doses reported by *Kito et al.*[26] For the *Thorax Vendor Default* protocol, $CTDI_{vol}$ was 12.6 mGy, which is comparable to the reported median of 9.6 mGy and falls within the overall range of 3.9-24.2 mGy reported for lung protocols across 42 CT simulators in the United Kingdom.[27]

Because accurate proton dose calculations demand higher image quality than image registration tasks typically performed before or during treatment, image quality was benchmarked using diagnostic CT phantoms and standards which were also consistent with vendor guidelines.[28] Technical metrics of image quality including high-contrast spatial resolution, low contrast performance, spatial integrity, CT number accuracy, and image uniformity were quantified using the ACR-464 phantom (Table 4). High-contrast spatial resolution exceeded the ACR minimum tolerance of 6 lp/cm, with 8 lp/cm observed by a board certified therapeutic medical physicist. This subjective assessment of spatial resolution was corroborated by MTF analysis, with a median 10% value of 7.1 lp/cm and a 50% MTF value of 4.4 lp/cm. These values compare favorably with those reported for recumbent CT simulators, ranging from 2.8-3.3 lp/cm at 50% MTF and 4.8-5.7 lp/cm at 10% MTF.[41]

While high-contrast spatial resolution met the ACR accreditation guidelines, future work may be warranted to implement additional reconstruction filters that trade some high-contrast spatial resolution for improved low-contrast performance. The upright CT system currently employs a high-pass Ram-Lak reconstruction kernel by default. Low contrast detectability is clinically important for delineation of soft tissue targets and organs-at-risk. In this study, the CNR ratio equaled 0.93 (range: 0.81-1.10), which narrowly failed to meet the ACR tolerance of greater than 1.0. Upon visual inspection, only the 25 mm diameter low-contrast object was resolved (Figure 2A). The smaller 6-mm diameter objects were not discernible. Visualization was partially limited by mild cupping artifacts attributed to image non-uniformity across the reconstruction FOV. Image uniformity measured 20.2 HU for the 20-cm diameter ACR-464 phantom, which was greater than the ≤7 HU ACR tolerance.

CT number accuracy was first evaluated for five material inserts. Measured values for water and acrylic fell within ACR-specified tolerances, while values for air, polyethylene, and bone showed minor deviations, with a maximum error of 27 HU. These small discrepancies are unlikely to meaningfully impact treatment planning and support the accurate conversion of CT number to mass density and SPR. Future work is warranted to refine beam hardening and scatter corrections to improve CT number accuracy to a level that meets ACR standards. When CT number measurements were conducted with a large Gammex Advanced Electron Density Phantom with additional inserts, upright CT number linearity was confirmed, with an $R^2$ value of 0.999 when compared to a reference recumbent CT scanner. This result supports accurate mapping to mass density and SPR across a broad range of clinically relevant tissue densities. Additionally, in-plane spatial integrity demonstrated a median error of 0.23 mm, well within the ≤1.0 mm tolerance recommended by the AAPM Task Group 66, further supporting the scanner's suitability for treatment planning.[31]

The feasibility of performing accurate photon and proton dose calculations on upright CT images was demonstrated in an anthropomorphic phantom study. Gamma analysis pass rates exceeded 97% using 2%/2mm criteria for a series of photon and proton dose distributions computed in homogenous and heterogenous regions such as lung and bone. Excellent agreement was maintained after imposing a stricter 1%/1mm criteria for photon plans (≥99.8%) as well as liver (99.9%) and spine (97.0%) proton plans. The stricter tolerance revealed some disagreement for the proton lung plan for which the gamma pass rate was 90.6% with the largest differences occurring in the low-density region. These findings support the clinical feasibility of upright CT for treatment planning, while also highlighting areas, particularly in highly heterogeneous low-density regions, where further refinement could be focused. Future work will include end-to-end validation with measurement-based comparisons once the upright CT system is integrated with a fixed proton beamline. A limitation of this study, consistent with current clinical practice, is that all proton dose calculations were based on single-energy CT and a corresponding CT number-to-SPR calibration. Incorporation of dual-energy CT derived SPRs may improve accuracy. Nevertheless, a recent NRG Oncology Survey reported that the majority (88%) of centers continue to rely on single-energy CT for proton therapy dose calculations rather than dual-energy CT.[42] Evaluation with dual-energy CT can be conducted as multi-energy capability becomes available for the upright CT scanner.

To the best of our knowledge, this study represents the first technical assessment of the upright CT system, which is undergoing a phase of rapid development. As a result, several current limitations of the system exist including a single calibrated imaging beam energy and only two reconstruction kernels. Furthermore, the system includes minimal raw projection data conditioning and calibration methods such as a beam hardening correction method and ring artifact correction. Advanced scatter correction, four-dimensional CT, and metal artifact reduction methods are not yet implemented on the system. Despite these limitations, the results of this study demonstrate the Marie system provides clinically acceptable imaging and dose calculation performance.

A limitation of our work is that all images were acquired over ~2 month period. Longitudinal image quality assessment studies are warranted to evaluate long term performance trends and stability. A thorough characterization of the patient positioner mechanical properties such as that performed by Sheng et al. is also warranted but beyond the scope of the current work focused on the upright CT imaging performance.[15] Finally, future work should compare the measured and vendor reported CT imaging dose which are expected to agree within 20% per guidelines from AAPM Task Group 66 and the ACR, which were not fully characterized at the time of the study.[28,31]

The technical characterization of the upright CT scanner presented here lays the foundation for acquiring high quality upright CT images in the treatment position and offers promise to advance image-guided and online adaptive proton therapy. Future work will focus on integrating the upright CT scanner with a fixed proton beamline, validating IGRT workflows, and systematically evaluating anatomical and dosimetric differences between upright and recumbent patient positioning to identify which disease sites will benefit.

## 5. CONCLUSION

This study demonstrated the feasibility of upright CT imaging for radiotherapy simulation and treatment planning. Image quality was thoroughly characterized against ACR accreditation standards. Proton and photon treatment planning and dose calculations showed accuracy comparable to those computed on a conventional recumbent CT system. These findings support the clinical viability of upright CT for radiotherapy applications.

**ACKNOWLEDGEMENTS**
Work reported was supported in part by the National Cancer Institute of the National Institutes of Health under award number R01HL153720 (principal investigator: Carri Glide-Hurst). The content is solely the responsibility of the authors and does not necessarily represent the official views of the National Institutes of Health. The authors would like to thank Brian Burger of Leo Cancer Care, Inc. for upright CT scanner technical support.

**CONFLICT OF INTEREST STATEMENT**
John Hayes and Carson Hoffman disclose employment by Leo Cancer Care, Inc. Jessica Miller has research collaboration funding with Siemens Healthineers. Carri Glide-Hurst reports research collaborations with RaySearch and Leo Cancer Care, Inc. pertaining to the work. Carri Glide-Hurst reports research collaborations with GE Healthcare and Medscint Inc outside the submitted work. Jordan Slagowski, Yuhao Yan, and Minglei Kang have no conflicts of interest to report.

**REFERENCES**

1. Rahim S, Korte J, Hardcastle N, Hegarty S, Kron T, Everitt S. Upright Radiation Therapy—A Historical Reflection and Opportunities for Future Applications. *Front Oncol*. 2020;10. doi:10.3389/fonc.2020.00213
2. Boisbouvier S, Boucaud A, Tanguy R, Grégoire V. Upright patient positioning for pelvic radiotherapy treatments. *Tech Innov Patient Support Radiat Oncol*. 2022;24:124-130. doi:10.1016/j.tipsro.2022.11.003
3. Gaito S, Aznar MC, Burnet NG, et al. Assessing Equity of Access to Proton Beam Therapy: A Literature Review. *Clin Oncol (R Coll Radiol)*. 2023;35(9):e528-e536. doi:10.1016/j.clon.2023.05.014
4. Mackie TR, Towe S, Schreuder N. Is upright radiotherapy medically and financially better? *AIP Conference Proceedings*. 2021;2348(1):020002. doi:10.1063/5.0051770
5. McCarroll RE, Beadle BM, Fullen D, et al. Reproducibility of patient setup in the seated treatment position: A novel treatment chair design. *J Appl Clin Med Phys*. 2017;18(1):223-229. doi:10.1002/acm2.12024
6. Yang J, Chu D, Dong L, Court LE. Advantages of simulating thoracic cancer patients in an upright position. *Pract Radiat Oncol*. 2014;4(1):e53-58. doi:10.1016/j.prro.2013.04.005
7. Yamada Y, Yamada M, Chubachi S, et al. Comparison of inspiratory and expiratory lung and lobe volumes among supine, standing, and sitting positions using conventional and upright CT. *Sci Rep*. 2020;10(1):16203. doi:10.1038/s41598-020-73240-8
8. Yamada Y, Yamada M, Chubachi S, et al. Comparison of inspiratory and expiratory airway volumes and luminal areas among standing, sitting, and supine positions using upright and conventional CT. *Sci Rep*. 2022;12(1):21315. doi:10.1038/s41598-022-25865-0
9. Marano J, Kissick MW, Underwood TSA, et al. Relative thoracic changes from supine to upright patient position: A proton collaborative group study. *Journal of Applied Clinical Medical Physics*. 2023;24(12):e14129. doi:10.1002/acm2.14129
10. Schreuder AN, Hsi WC, Greenhalgh J, et al. Anatomical changes in the male pelvis between the supine and upright positions-A feasibility study for prostate treatments in the upright position. *J Appl Clin Med Phys*. 2023;24(11):e14099. doi:10.1002/acm2.14099
11. Reiff JE, Werner-Wasik M, Valicenti RK, Huq MS. Changes in the size and location of kidneys from the supine to standing positions and the implications for block placement during total body irradiation. *Int J Radiat Oncol Biol Phys*. 1999;45(2):447-449. doi:10.1016/s0360-3016(99)00208-4
12. Mohiuddin MM, Zhang B, Tkaczuk K, Khakpour N. Upright, standing technique for breast radiation treatment in the morbidly-obese patient. *Breast J*. 2010;16(4):448-450. doi:10.1111/j.1524-4741.2010.00932.x
13. Miller RW, Raubitschek AA, Harrington FS, van de Geijn J, Ovadia J, Glatstein E. An isocentric chair for the simulation and treatment of radiation therapy patients. *Int J Radiat Oncol Biol Phys*. 1991;21(2):469-473. doi:10.1016/0360-3016(91)90798-9
14. Zhang X, Hsi WC, Yang F, et al. Development of an isocentric rotating chair positioner to treat patients of head and neck cancer at upright seated position with multiple nonplanar fields in a fixed carbon-ion beamline. *Med Phys*. 2020;47(6):2450-2460. doi:10.1002/mp.14115
15. Sheng Y, Sun J, Wang W, et al. Performance of a 6D Treatment Chair for Patient Positioning in an Upright Posture for Fixed Ion Beam Lines. *Front Oncol*. 2020;10. doi:10.3389/fonc.2020.00122
16. Korte JC, Wright M, Krishnan PG, et al. A radiation therapy platform to enable upright cone beam computed tomography and future upright treatment on existing photon therapy machines. *Med Phys*. 2025;52(2):1133-1145. doi:10.1002/mp.17523

17. Feldman J, Pryanichnikov A, Achkienasi A, et al. Commissioning of a novel gantry-less proton therapy system. *Front Oncol*. 2024;14:1417393. doi:10.3389/fonc.2024.1417393
18. Volz L, Korte J, Martire MC, et al. Opportunities and challenges of upright patient positioning in radiotherapy. *Phys Med Biol*. 2024;69(18):18TR02. doi:10.1088/1361-6560/ad70ee
19. Volz L, Sheng Y, Durante M, Graeff C. Considerations for Upright Particle Therapy Patient Positioning and Associated Image Guidance. *Front Oncol*. 2022;12. doi:10.3389/fonc.2022.930850
20. Jinzaki M, Yamada Y, Nagura T, et al. Development of Upright Computed Tomography With Area Detector for Whole-Body Scans: Phantom Study, Efficacy on Workflow, Effect of Gravity on Human Body, and Potential Clinical Impact. *Invest Radiol*. 2020;55(2):73-83. doi:10.1097/RLI.0000000000000603
21. Fukuoka R, Yamada Y, Kataoka M, et al. Estimating right atrial pressure using upright computed tomography in patients with heart failure. *Eur Radiol*. 2023;33(6):4073-4081. doi:10.1007/s00330-022-09360-8
22. Kosugi K, Yamada Y, Yamada M, et al. Posture-induced changes in the vessels of the head and neck: evaluation using conventional supine CT and upright CT. *Sci Rep*. 2020;10(1):16623. doi:10.1038/s41598-020-73658-0
23. Norimatsu T, Nakahara T, Yamada Y, et al. Anatomical cardiac and electrocardiographic axes correlate in both upright and supine positions: an upright/supine CT study. *Sci Rep*. 2023;13(1):18170. doi:10.1038/s41598-023-45528-y
24. Feldman J, Pryanichnikov A, Shwartz D, et al. Study of upright patient positioning reproducibility in image-guided proton therapy for head and neck cancers. *Radiother Oncol*. 2024;201:110572. doi:10.1016/j.radonc.2024.110572
25. Kissick MW, Panaino C, Criscuolo A, et al. Calculation method for novel upright CT scanner isodose curves. *Journal of Applied Clinical Medical Physics*. 2024;25(7):e14377. doi:10.1002/acm2.14377
26. Kito S, Suda Y, Tanabe S, et al. Radiological imaging protection: a study on imaging dose used while planning computed tomography for external radiotherapy in Japan. *J Radiat Res*. 2023;65(2):159-167. doi:10.1093/jrr/rrad098
27. Wood TJ, Davis AT, Earley J, et al. IPEM topical report: the first UK survey of dose indices from radiotherapy treatment planning computed tomography scans for adult patients. *Phys Med Biol*. 2018;63(18):185008. doi:10.1088/1361-6560/aacc87
28. Chad Dillon M, William Breeden M III, Jessica Clements M, et al. ACR Computed Tomography Quality Control Manual. https://www.acr.org/-/media/ACR/Files/Clinical-Resources/QC-Manuals/CT_QCManual.pdf
29. Friedman SN, Fung GSK, Siewerdsen JH, Tsui BMW. A simple approach to measure computed tomography (CT) modulation transfer function (MTF) and noise-power spectrum (NPS) using the American College of Radiology (ACR) accreditation phantom. *Med Phys*. 2013;40(5):051907. doi:10.1118/1.4800795
30. Hobson MA, Soisson ET, Davis SD, Parker W. Using the ACR CT accreditation phantom for routine image quality assurance on both CT and CBCT imaging systems in a radiotherapy environment. *J Appl Clin Med Phys*. 2014;15(4):226-239. doi:10.1120/jacmp.v15i4.4835
31. Mutic S, Palta JR, Butker EK, et al. Quality assurance for computed-tomography simulators and the computed-tomography-simulation process: Report of the AAPM Radiation Therapy Committee Task Group No. 66. *Medical Physics*. 2003;30(10):2762-2792. doi:10.1118/1.1609271
32. Peters N, Trier Taasti V, Ackermann B, et al. Consensus guide on CT-based prediction of stopping-power ratio using a Hounsfield look-up table for proton therapy. *Radiother Oncol*. 2023;184:109675. doi:10.1016/j.radonc.2023.109675
33. "Cheese" Phantoms - pylinac 3.30.0 documentation. Accessed December 26, 2024. https://pylinac.readthedocs.io/en/latest/cheese.html

34. Redmond KJ, Lo SS, Soltys SG, et al. Consensus guidelines for postoperative stereotactic body radiation therapy for spinal metastases: results of an international survey. *J Neurosurg Spine*. 2017;26(3):299-306. doi:10.3171/2016.8.SPINE16121
35. Nguyen QN, Ly NB, Komaki R, et al. Long-term outcomes after proton therapy, with concurrent chemotherapy, for stage II-III inoperable non-small cell lung cancer. *Radiother Oncol*. 2015;115(3):367-372. doi:10.1016/j.radonc.2015.05.014
36. Benedict SH, Yenice KM, Followill D, et al. Stereotactic body radiation therapy: the report of AAPM Task Group 101. *Med Phys*. 2010;37(8):4078-4101. doi:10.1118/1.3438081
37. Hong TS, Wo JY, Borger DR, et al. Phase II Study of Proton-Based Stereotactic Body Radiation Therapy for Liver Metastases: Importance of Tumor Genotype. *J Natl Cancer Inst*. 2017;109(9). doi:10.1093/jnci/djx031
38. Low DA, Harms WB, Mutic S, Purdy JA. A technique for the quantitative evaluation of dose distributions. *Med Phys*. 1998;25(5):656-661. doi:10.1118/1.598248
39. Biggs S, Jennings M, Swerdloff S, et al. PyMedPhys: A community effort to develop an open, Python-based standard library for medical physics applications. *Journal of Open Source Software*. 2022;7(78):4555. doi:10.21105/joss.04555
40. Miften M, Olch A, Mihailidis D, et al. Tolerance limits and methodologies for IMRT measurement-based verification QA : *Recommendations of AAPM Task Group No. 218* . *Medical Physics*. 2018;45(4). doi:10.1002/mp.12810
41. Tomic N, Papaconstadopoulos P, Aldelaijan S, Rajala J, Seuntjens J, Devic S. Image quality for radiotherapy CT simulators with different scanner bore size. *Phys Med*. 2018;45:65-71. doi:10.1016/j.ejmp.2017.11.017
42. Lin L, Taylor PA, Shen J, et al. NRG Oncology Survey of Monte Carlo Dose Calculation Use in US Proton Therapy Centers. *Int J Part Ther*. 2021;8(2):73-81. doi:10.14338/IJPT-D-21-00004